# A Cut-off Phenomenon in Location Based Random Access Games with Imperfect Information

| Hazer Inaltekin | Mung Chiang | H. Vincent Poor |
|---|---|---|
| Department of Electrical Engineering | Department of Electrical Engineering | Department of Electrical Engineering |
| Princeton University | Princeton University | Princeton University |
| Princeton, NJ | Princeton, NJ | Princeton, NJ |
| hinaltek@princeton.edu | chiangm@princeton.edu | poor@princeton.edu |

## ABSTRACT

This paper analyzes the behavior of selfish transmitters under imperfect location information. The scenario considered is that of a wireless network consisting of selfish nodes that are randomly distributed over the network domain according to a known probability distribution, and that are interested in communicating with a common sink node using common radio resources. In this scenario, the wireless nodes do not know the exact locations of their competitors but rather have belief distributions about these locations. Firstly, properties of the packet success probability curve as a function of the node-sink separation are obtained for such networks. Secondly, a monotonicity property for the best-response strategies of selfish nodes is identified. That is, for any given strategies of competitors of a node, there exists a critical node-sink separation for this node such that its best-response is to transmit when its distance to the sink node is smaller than this critical threshold, and to back off otherwise. Finally, necessary and sufficient conditions for a given strategy profile to be a Nash equilibrium are provided.

## Categories and Subject Descriptors

C.2 [**Computer-Communication Networks**]: Network Architecture and Design—*Wireless communication*

*The research was supported by the U.S. National Science Foundation under Grants ANI-03-38807 and CNS-06-25637.

## 1. INTRODUCTION

### 1.1 Background and Motivation

The design of effective, distributed and scalable protocols is complicated when the application environment is hostile. If we allow for the possibility of nodes being captured and modified by malicious agents, the design problem is particularly difficult. One way of dealing with these issues and to have a sense of what kind of network behavior arises is to invest all the decision-making burden in individual network nodes. Nodes selfishly decide what to do by sensing their local environment with the aim of maximizing their own utilities. Such an approach by its very nature results in distributed and scalable network control and management. *Game theory* provides the necessary mathematical tools to analyze the emergent behavior of networks of such selfish agents. For this reason, game theory will lie at the heart of design methodologies for next-generation complex and self-organizing wireless networks.

In this paper, we consider a network of wireless *selfish* nodes trying to independently maximize their utility functions. We assume that these selfish nodes are willing to communicate with a sink node, and are randomly distributed over the network domain according to a known probability distribution. As opposed to much of the existing game theoretic work in the context of wireless networking (e.g., [1], [2] and [3]), we do not assume perfect information about locations of wireless nodes. The first motivating reason for imperfect location information is that distributing node location information to all network nodes is a prohibitive task in wireless networks containing large numbers of nodes. The second motivating reason is that malicious nodes tend to report false location information. Therefore, it becomes imperative to understand the behavior of selfish nodes under *imperfect location information* about others.

To this end, we assume that node locations are determined by nature according to a spatial point process. The distribution of this point process is known by all nodes, but a node remains unsure of the exact realizations of its opponents' locations. Hence, each node tries to maximize its expected utility by taking uncertainties about other nodes' locations into account. Our results can be summarized as follows. Let $\mathcal{N}$ be the set of all nodes wishing to communicate with a sink node. For any given strategy profile $\mathbf{s}_{-i} = (s_j)_{j \in \mathcal{N}-\{i\}}$ of other nodes, we prove that packet success probability of node $i$ is a non-increasing continuous function of its distance to the sink node. This property enables us to prove a monotonicity property of best-response strategies of nodes. In particular, we show that for any given strategy profile $\mathbf{s}_{-i} = (s_j)_{j \in \mathcal{N}-\{i\}}$ of other nodes, there exists a critical threshold $d_i^*$ of node $i$ such that node $i$'s best-response strategy is to transmit when its distance to the sink is smaller than $d_i^*$, and to back off otherwise. We finally give necessary and sufficient conditions for a strategy profile to be a Nash equilibrium strategy profile.

## 1.2 Related Work

Recently, there has been intensified research interest in the game theoretic modeling of wireless networks at the protocol level; see e.g., [1], [2] and [3]. This line of research is mostly motivated by practical considerations revolving around the design, deployment, control and management of complex wireless networks of the near future. Designing a wireless network containing a large number of nodes imposes challenging tasks like proposing scalable network protocols and algorithms, and making sure that every node runs them without any deviation.

In [1] and [4], the authors model slotted ALOHA with multi-packet reception in the presence of selfish users by using repeated games. They consider only the homogeneous case where all nodes are identical, and prove the existence of a symmetric Nash equilibrium. The paper [2] addresses the selfish behavior in the CSMA/CA MAC protocol. The authors of [2] derive the conditions for the optimal functioning of the selfish transmitters in CSMA/CA. They also propose an algorithm that guides selfish nodes to a Pareto-efficient Nash equilibrium. In [3] and [5], a one-shot random access game for wireless networks is introduced, and an in-depth analysis of the channel throughput at Nash equilibria is given. The authors also provide a detailed analysis of the asymptotic properties of the game as the number of selfish transmitters increases without bound.

In all of the above work, the common assumption is that every player of the game knows the state (e.g., positions) of other players perfectly. In contrast to them, we take a different approach in this paper. Our selfish nodes are randomly distributed over the network domain, and do not know the exact positions of their opponents. Only the probability distribution according to which node locations are drawn is known and common information to all nodes. Selfish nodes maximize their expected utilities by taking uncertainties in node locations into account. This work analyzes the behavior of such selfish nodes under *imperfect location information* when they communicate with a common sink node.

## 2. NETWORK MODEL AND NODE STRATEGIES

### 2.1 Network Model

We consider a disk shaped network domain with radius $R > 0$. $n$ nodes are randomly distributed over the network domain according to a spatial point process. We assume that there is a common sink node located at the center of the network domain with which all nodes wish to communicate. This common node can be thought of as being the base station in the context of cellular networks, or as being the fusion center in the context of wireless sensor networks.

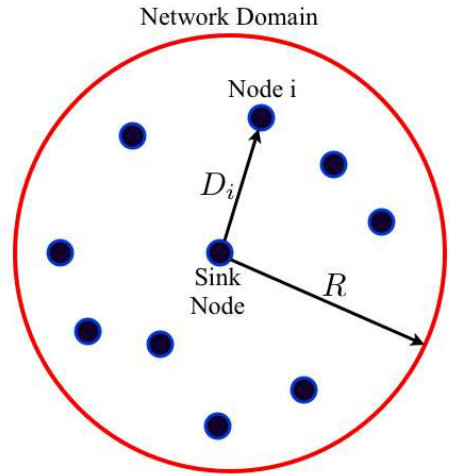

**Figure 1: A particular realization of the network.**

We let $D_i$ represent the distance of the $i$th node, $i \in \mathcal{N} = \{1, 2, \cdots, n\}$, to the sink node. A particular realization of $D_i$ will be represented by $d_i$. When there is no confusion, we also sometimes denote the node-sink separation by $d$ without any indexes. We assume that $\{D_i\}_{i=1}^n$ form an independent and identically distributed set of random variables with a common probability distribution $\mu$. $\mu$ is common knowledge to all nodes, and it is assumed to be an *absolutely continuous* probability measure with respect to the Lebesgue measure $\lambda$ on $[0, R]$. We write this property as $\mu \ll \lambda$. A particular realization of this network is depicted in Fig. 1.

We assume that a transmitted packet is successfully received by the sink node if it is the packet received with the highest signal power. We further assume that there is no power control algorithm employed at the physical layer, and all nodes transmit their data packets at the same power level. The transmitted signal power monotonically decreases as a function of the distance. Therefore, a packet from node $i$ will be successful if node $i$ is the closest transmitting node to the sink node.

The cost of unsuccessful transmission for node $i$ is $c_i \in (0, \infty)$. If a transmission is successful, the node which transmitted its packet successfully receives utility 1.

### 2.2 Node Strategies

We define the strategy of a node as a function from all possible values of node-sink separation to the two point set $\{0, 1\}$, where 0 means back off and 1 means transmission. Therefore, the strategy of a node determines whether it transmits or backs off for any given node-sink separation. We formally define the strategy of node $i$ as follows.

DEFINITION 1. *A strategy of node $i$ is a function*

$$s_i : [0, R] \to \{0, 1\},$$

*which determines whether to transmit or not for any given node-sink separation $d \in [0, R]$.*

A strategy profile $\mathbf{s} = (s_1, s_2, \cdots, s_n)$ is the vector of strategies of nodes. When we write $\mathbf{s}_{-i} = (s_j)_{j \in \mathcal{N}-\{i\}}$, we mean the strategy profile containing strategies of all nodes except node $i$.

# 3. BEST RESPONSE STRATEGIES AND NASH EQUILIBRIA

## 3.1 Properties of Packet Success Probability

We will first look at some basic properties of the packet success probability for any given arbitrary strategy profile **s**. Let $E_i(d)$ be the event that the packet of $i$th transmitter is successful when the distance between the sink node and the $i$th node is $d$. We let

$$g_i(d) = \mathbb{P}(E_i(d)). \tag{1}$$

PROPERTY 1. $g_i(0) = 1$ for any given strategy profile $\mathbf{s}_{-i} = (s_j)_{j \in \mathcal{N} - \{i\}}$ of other nodes.

PROOF.
$$\begin{aligned}
g_i(d) &= \mathbb{P}\{\forall j \in \mathcal{N} - \{i\} : s_j(D_j) = 0 \text{ or } D_j > d\} \\
&= \prod_{j \in \mathcal{N} - \{i\}} \mu\{d_j \in [0, R] : s_j(d_j) = 0 \text{ or } d_j > d\} \\
&= \prod_{j \in \mathcal{N} - \{i\}} \mu\left((d, R] \cup \{d_j \in [0, R] : s_j(d_j) = 0\}\right) \\
&\geq \mu((d, R])^{n-1}.
\end{aligned}$$

Since $\mu \ll \lambda$, $\mu((0, R]) = 1$. Therefore, $g_i(0) = 1$. □

DEFINITION 2. Given the strategy $s_j$ of node $j$, its subjective transmission probability (from the perspective of other nodes) is equal to

$$b_j = \mu\{d_j \in [0, R] : s_j(d_j) = 1\}. \tag{2}$$

Since the belief distribution about the location of node $j$ is the same for all nodes, $b_j$ becomes the same for all of them. In cases where different network nodes have different belief distributions about the location of node $j$, its subjective transmission probability changes from node to node.

PROPERTY 2. For any given strategy profile $\mathbf{s}_{-i} = (s_j)_{j \in \mathcal{N} - \{i\}}$ of other nodes, $g_i(d)$ is a non-increasing function of $d$. Moreover, if there exists at least one node in $\mathcal{N} - \{i\}$ having positive subjective transmission probability, then $g_i(R) < 1$.

PROOF. Let
$$q_j(d) = \mu\left((d, R] \cup \{d_j \in [0, R] : s_j(d_j) = 0\}\right).$$

Then, $g_i(d) = \prod_{j \in \mathcal{N} - \{i\}} q_j(d)$. It is easy to see that $q_j(d) \geq q_j(d')$ for all $d' \geq d$. Thus, $g_i(d)$ is a non-increasing function of $d$. Now, assume that there exists a node $j \in \mathcal{N} - \{i\}$ such that $b_j > 0$. Then,

$$\begin{aligned}
q_j(R) &= \mu(\{R\} \cup \{d_j \in [0, R] : s_j(d_j) = 0\}) \\
&\stackrel{(a)}{=} \mu\{d_j \in [0, R] : s_j(d_j) = 0\} \\
&= 1 - b_j \\
&< 1,
\end{aligned}$$

where (a) follows from $\mu \ll \lambda$. $q_j(R)$ being smaller than 1 for some $j \in \mathcal{N} - \{i\}$ also implies that $g_i(R) < 1$. □

PROPERTY 3. Consider an interval $(a, b) \subseteq [0, R]$, and let $\mathbf{s}_{-i} = (s_j)_{j \in \mathcal{N} - \{i\}}$ be the strategy profile of other nodes.

- If $s_j = 0$ on $(a, b)$ for all $j \in \mathcal{N} - \{i\}$, then $g_i$ is constant on $(a, b)$.
- If $\lambda \ll \mu$ and there exists $j \in \mathcal{N} - \{i\}$ such that $s_j = 1$ on $(a, b)$, then $g_i$ is strictly decreasing on $(a, b)$.

PROOF. Let
$$q_j(d) = \mu\left((d, R] \cup \{d_j \in [0, R] : s_j(d_j) = 0\}\right).$$

We will show that $q_j$ is constant on $(a, b)$ if $s_j = 0$ on $(a, b)$. Assume $s_j = 0$ on $(a, b)$ and take any $\alpha$ and $\beta$ in $(a, b)$. Without loss of generality, assume that $\alpha \leq \beta$. Then,

$$\begin{aligned}
q_j(\alpha) &- q_j(\beta) \\
&= \mu\left((\alpha, \beta] - \{d_j \in [0, R] : s_j(d_j) = 0\}\right) \\
&= \mu(\emptyset) = 0.
\end{aligned}$$

This proves that $q_j$ is constant on $(a, b)$ if $s_j = 0$ on $(a, b)$. Since $g_i(d) = \prod_{j \in \mathcal{N} - \{i\}} q_j(d)$, $g_i$ also becomes constant on $(a, b)$ if $s_j = 0$ for all $j \in \mathcal{N} - \{i\}$.

To prove the second claim, assume that $s_j = 1$ on $(a, b)$ for some $j \in \mathcal{N} - \{i\}$. Then, for any $\alpha$ and $\beta$ in $(a, b)$ with $\alpha < \beta$, we have

$$\begin{aligned}
q_j(\alpha) &- q_j(\beta) \\
&= \mu\left((\alpha, \beta] - \{d_j \in [0, R] : s_j(d_j) = 0\}\right) \\
&= \mu((\alpha, \beta]) > 0,
\end{aligned}$$

where the last inequality follows from $\lambda \ll \mu$. This proves that $q_j$ is strictly decreasing on $(a, b)$ for at least one $j \in \mathcal{N} - \{i\}$. Since $g_i(d) = \prod_{j \in \mathcal{N} - \{i\}} q_j(d)$ and $q_j(d)$'s are non-increasing with at least one of them strictly decreasing $(a, b)$, we also have $g_i(d)$ is strictly decreasing on $(a, b)$. □

DEFINITION 3. The density of $\mu$ is its Radon-Nikodym derivative $f$ with respect to $\lambda$.

If $f$ is the density of $\mu$, the following holds for any measurable subset $E$ of $[0, R]$:

$$\mu(E) = \int_E f(x) d\lambda(x).$$

PROPERTY 4. For any given strategy profile $\mathbf{s}_{-i} = (s_j)_{j \in \mathcal{N} - \{i\}}$ of other nodes, $g_i(d)$ is a continuous function of $d$. Moreover, it $\mu$ has a bounded density, it is Hölder continuous of order $n - 1$.

PROOF. Let again
$$q_j(d) = \mu\left((d, R] \cup \{d_j \in [0, R] : s_j(d_j) = 0\}\right).$$

For any given $\delta > 0$, choose $\epsilon > 0$ small enough that $\mu((d, d + \epsilon])$ is smaller than $\delta$. This is possible since $\mu \ll \lambda$ (see [6]). Then,

$$\begin{aligned}
|q_j(d) - q_j(d + \epsilon)| &= \mu\left((d, d + \epsilon] - \{d_j \in [0, R] : s_j(d_j) = 0\}\right) \\
&\leq \mu((d, d + \epsilon]) \\
&\leq \delta.
\end{aligned}$$

This proves the continuity of $q_j(d)$ as a function of $d$ for all $j \in \mathcal{N} - \{i\}$. Since $g_i(d) = \prod_{j \in \mathcal{N} - \{i\}} q_j(d)$, $g_i(d)$ is also continuous as a function of $d$.

Now, assume that $\mu$ has a bounded density $f$. Let $K = \sup_{x\in[0,R]} f(x) < \infty$. Then,

$$\begin{aligned} |q_j(d) - q_j(d+\epsilon)| &\leq \mu((d, d+\epsilon]) \\ &= \int_d^{d+\epsilon} f(x) d\lambda(x) \\ &\leq \epsilon K. \end{aligned}$$

Therefore, $|g_i(d) - g_i(d+\epsilon)| \leq \epsilon^{n-1} K^{n-1}$. $\square$

The packet success probability curves are illustrated more concretely in following examples where there are two nodes uniformly distributed over the network domain. In these examples, we focus on $g_1(d)$ and vary the strategy of node 2.

EXAMPLE 1. *Let $s_2(d) = 1$ for $d \in \left[0, \frac{R}{2}\right]$ and $s_2(d) = 0$ for $d \in \left(\frac{R}{2}, R\right]$. Then,*

$$\begin{aligned} g_1(d) &= \mu\left((d, R] \cup \left(\frac{R}{2}, R\right]\right) \\ &= \begin{cases} \frac{R^2 - d^2}{R^2} & \text{if } 0 \leq d < \frac{R}{2}, \\ \frac{3}{4} & \text{if } \frac{R}{2} \leq d \leq R \end{cases}. \end{aligned}$$

EXAMPLE 2. *Let $s_2(d) = 0$ for $d \in \left[0, \frac{R}{2}\right)$ and $s_2(d) = 1$ for $d \in \left[\frac{R}{2}, R\right]$. Then,*

$$\begin{aligned} g_1(d) &= \mu\left((d, R] \cup \left[0, \frac{R}{2}\right)\right) \\ &= \begin{cases} 1 & \text{if } 0 \leq d < \frac{R}{2}, \\ \frac{1}{4} + \frac{R^2 - d^2}{R^2} & \text{if } \frac{R}{2} \leq d \leq R \end{cases}. \end{aligned}$$

EXAMPLE 3. *Let $s_2(d) = 0$ for $d \in \left(\frac{R}{3}, \frac{2R}{3}\right)$, and 1 otherwise. Then,*

$$\begin{aligned} g_1(d) &= \mu\left((d, R] \cup \left(\frac{R}{3}, \frac{2R}{3}\right)\right) \\ &= \begin{cases} \frac{R^2 - d^2}{R^2} & \text{if } 0 \leq d < \frac{R}{3}, \\ \frac{8}{9} & \text{if } \frac{R}{3} \leq d < \frac{2R}{3}, \\ \frac{1}{3} + \frac{R^2 - d^2}{R^2} & \text{if } \frac{2R}{3} \leq d \leq R \end{cases}. \end{aligned}$$

EXAMPLE 4. *Let $s_2(d) = 1$ for $d \in \left[\frac{R}{3}, \frac{2R}{3}\right]$, and 0 otherwise. Then,*

$$\begin{aligned} g_1(d) &= \mu\left((d, R] \cup \left[0, \frac{R}{3}\right) \cup \left(\frac{2R}{3}, R\right]\right) \\ &= \begin{cases} 1 & \text{if } 0 \leq d < \frac{R}{3}, \\ \frac{1}{9} + \frac{R^2 - d^2}{R^2} & \text{if } \frac{R}{3} \leq d < \frac{2R}{3}, \\ \frac{2}{3} & \text{if } \frac{2R}{3} \leq d \leq R \end{cases}. \end{aligned}$$

Node 2 strategies in above examples and the corresponding $g_1(d)$ for these examples are shown in Fig. 2 as a function of $d$ by setting $R = 12$. The properties of the packet success probability listed above can be observed more concretely in this figure.

## 3.2 Best Response Strategies

We now investigate the behavior of the best-response strategy $s_i^*$ of node $i$ given the strategies $\mathbf{s}_{-i} = (s_j)_{j \in \mathcal{N}-\{i\}}$ of other nodes. To formally define best-response strategies of nodes, we let $\mathbb{E}[u_i|T](d)$ be the expected utility of node $i$ given that it transmits and its distance to the sink node is equal to $d$. $\mathbb{E}[u_i|T](d)$ also depends on $\mathbf{s}_{-i}$ but we do not show this dependence explicitly to keep the notation as simple as possible. Then, a selfish node transmits if its expected utility given that it transmits is greater than zero. On the other hand, it defers from transmission if its expected utility given that it transmits is smaller than zero. We break the ties at 0 by assuming that node $i$ backs off when its expected utility given that it transmits is equal to zero. We formally define the best-response strategy of node $i$ as follows.

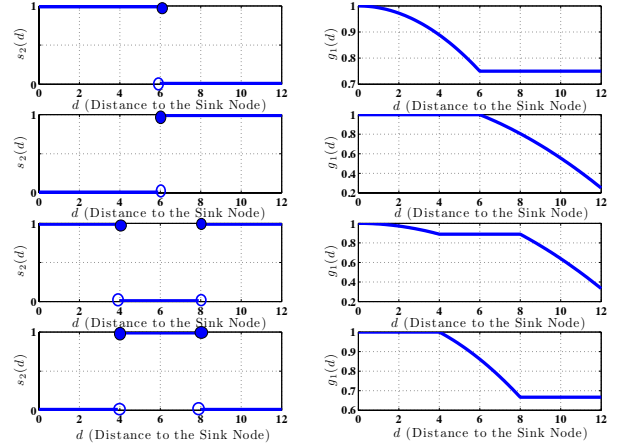

**Figure 2: Change of packet success probability $g_1(d)$ as a function of $d$ when there two nodes uniformly distributed over the network domain. $g_1(d)$ is plotted for four different strategies of node 2.**

DEFINITION 4. *For a given strategy profile $\mathbf{s}_{-i} = (s_j)_{j \in \mathcal{N}-\{i\}}$ of other nodes, let*

$$\mathcal{B}_i = \{d \in [0, R] : \mathbb{E}[u_i|T](d) > 0\}.$$

*Then, we call $\mathcal{B}_i$ the* best-response set *of node $i$ given $\mathbf{s}_{-i}$. The* best-response strategy *$s_i^*$ of node $i$ given $\mathbf{s}_{-i}$ is to transmit for all $d_i \in \mathcal{B}_i$, and to back off otherwise.*

DEFINITION 5. *We say the strategy of node $i$ is a monotonically decreasing strategy if there exists a cut-off point $\tilde{d}_i$ such that node $i$ transmits when its distance to the sink is smaller than $\tilde{d}_i$, and backs off when its distance to the sink is greater than or equal to $\tilde{d}_i$. A strategy profile $\mathbf{s} = (s_i)_{i \in \mathcal{N}}$ is a monotonically decreasing strategy profile if $s_i$ is a monotonically decreasing strategy for all $i \in \mathcal{N}$.*

$\mathbb{E}[u_i|T](d)$ can be written explicitly as follows:

$$\begin{aligned} \mathbb{E}[u_i|T](d) &= g_i(d) - c_i(1 - g_i(d)) \\ &= (1 + c_i)g_i(d) - c_i. \end{aligned}$$

Therefore, the properties proven for $g_i(d)$ above continue to hold for $\mathbb{E}[u_i|T](d)$. In particular, $\mathbb{E}[u_i|T](d)$ is a non-increasing and continuous function of $d$ by Property 2 and Property 4. This implies that whenever $\mathbb{E}[u_i|T](d') > 0$ for some $d' \in [0, R]$, $\mathbb{E}[u_i|T](d) > 0$ for all $d \leq d'$. Similarly, whenever $\mathbb{E}[u_i|T](d') \leq 0$ for some $d' \in [0, R]$, $\mathbb{E}[u_i|T](d) \leq 0$ for all $d \geq d'$.

The above observations enable us to characterize best-response strategies of nodes by means of least upper bounds of their best-response sets. In particular, we will show that for any given strategy profile $\mathbf{s}_{-i} = (s_j)_{j \in \mathcal{N}-\{i\}}$ of other nodes, there exists a critical node-sink separation $d_i^*$ such that if node $i$'s distance to the sink node is smaller than $d_i^*$, then node $i$ transmits. Otherwise, it backs off. Moreover, $d_i^*$ is equal to the least upper bound of $\mathcal{B}_i$. Therefore, the best-response strategy of node $i$ against any $\mathbf{s}_{-i}$ is a monotonically decreasing strategy whose cut-off point is equal to $d_i^*$. These claims are formally stated in Theorem 1.

THEOREM 1. *Let $\mathbf{s}_{-i} = (s_j)_{j \in \mathcal{N}-\{i\}}$ be a strategy profile of other nodes, $d$ be the node-sink separation and*

$$d_i^* = \sup \mathcal{B}_i.$$

*Then, the following hold:*

- $d_i^* > 0$.

- *If $\mathbb{E}[u_i|T](R) > 0$, then $d_i^* = R$ and node $i$ transmits for all $d \in [0, R]$.*

- *If $\mathbb{E}[u_i|T](R) = 0$ and $d_i^* = R$, then node $i$ transmits for all $d \in [0, R)$, and backs off for $d = R$.*

- *If $d_i^* \in (0, R)$, then $\mathbb{E}[u_i|T](d_i^*) = 0$ and node $i$ transmits for all $d \in [0, d_i^*)$ and backs off for all $d \in [d_i^*, R]$.*

PROOF. The first item follows from the continuity of $\mathbb{E}[u_i|T](d)$ and the fact that $\mathbb{E}[u_i|T](0) = 1$. Let us focus on the second item. If $\mathbb{E}[u_i|T](R) > 0$, then $\mathbb{E}[u_i|T](d) > 0$ for all $d \in [0, R]$ since $\mathbb{E}[u_i|T](d)$ is a non-increasing function of $d$. This proves the second item.

For the third item, we will show that $\mathbb{E}[u_i|T](d) > 0$ for all $d \in [0, R)$. Suppose this is not correct. Then, there exists $d' < R$ such that $\mathbb{E}[u_i|T](d') \leq 0$. Since $\mathbb{E}[u_i|T](0) = 1$, we have $\mathbb{E}[u_i|T](d'') = 0$ for some $d'' < R$ by the intermediate value theorem. Let

$$d_* = \inf \{d \in [0, R] : \mathbb{E}[u_i|T](d) = 0\}. \quad (3)$$

$d_*$ is the first time $\mathbb{E}[u_i|T](d)$ hits 0. Since $\mathbb{E}[u_i|T](d)$ is a non-increasing continuous function starting at 1 when $d = 0$ and crossing 0 for some positive $d$ smaller than $R$, we have $d_i^* = d_* < R$, which is a contradiction.

To prove the fourth item, first observe that $\mathbb{E}[u_i|T](d) \leq 0$ for all $d \in [d_i^*, R]$ by using the definition of $d_i^*$ and the continuity of $\mathbb{E}[u_i|T](d)$. Since $\mathbb{E}[u_i|T](0) = 1$, there exists a $d' \leq d_i^*$ such that $\mathbb{E}[u_i|T](d') = 0$ by the intermediate value theorem. We define $d_*$ as in (3). Since $\mathbb{E}[u_i|T](d)$ is a non-increasing continuous function starting at 1 when $d = 0$ and crossing 0 for some positive $d$, $d_i^* = d_*$. We have $\mathbb{E}[u_i|T](d_i^*) = 0$ by using $d_i^* = d_*$ and the continuity of $\mathbb{E}[u_i|T](d)$. Observe that $\mathbb{E}[u_i|T](d) > 0$ for all $d \in [0, d_i^*)$ since $d_i^*$ is the first time that $\mathbb{E}[u_i|T](d)$ hits 0. As a result, node $i$ transmits for all $d \in [0, d_i^*)$, and backs off for all $d \in [d_i^*, R]$. □

### 3.3 Properties of Nash Equilibrium Strategy Profiles

We now focus on Nash equilibria of the random access game with imperfect location information. We formally define a Nash equilibrium strategy profile as follows.

DEFINITION 6. *A strategy profile $\mathbf{s}^* = (s_i^*)_{i \in \mathcal{N}}$ is a Nash equilibrium of the random access game with imperfect location information if and only if for all $i \in \mathcal{N}$, $s_i^*$ is the best-response strategy of node $i$ given strategies $\mathbf{s}_{-i}^* = (s_j^*)_{j \in \mathcal{N}-\{i\}}$ of other nodes.*

Due to the monotonicity property of best-response strategies proved in Theorem 1, a Nash equilibrium strategy profile becomes a monotonically decreasing strategy profile, and we can characterize Nash equilibrium strategies by means of critical node-sink separations. To this end, for any given Nash equilibrium strategy profile $\mathbf{s}^* = (s_i^*)_{i \in \mathcal{N}}$, we let

$$d_i^* = \sup \mathcal{B}_i. \quad (4)$$

It follows from Theorem 1 and the definition of Nash equilibrium that at a Nash equilibrium, node $i$ transmits until $d_i^*$ and backs off after $d_i^*$, and this is correct for all $i \in \mathcal{N}$. In the rest of the paper, we will analyze the properties of critical node-sink separations. The following theorem shows that there cannot be two different nodes transmitting for all node-sink separations at a Nash equilibrium.

THEOREM 2. *At a Nash equilibrium, there cannot be two different nodes $i$ and $j$ in $\mathcal{N}$ having $d_i^* = d_j^* = R$.*

PROOF. Suppose there are two different nodes $i$ and $j$ in $\mathcal{N}$ having $d_i^* = d_j^* = R$. Let $d$ be the distance of node $i$ to the sink node. Then, packets from node $i$ fails if $D_j < d$. Therefore, we have the following upper bound on the packet success probability of node $i$ when its separation from the sink node is equal to $d$:

$$g_i(d) \leq 1 - \mathbb{P}\{D_j < d\}.$$

Thus, $g_i(d)$ goes to zero as $d$ increases to $R$. As a result,

$$\mathbb{E}[u_i|T](d) = (1 + c_i)g_i(d) - c_i$$

eventually becomes smaller than 0, and stays below 0 for all $d$ close enough to $R$, which contradicts $d_i^* = R$. □

In the next theorem, we establish necessary conditions for nodes' strategies at Nash equilibria.

THEOREM 3. *Let $\mathbf{s}^* = (s_i^*)_{i \in \mathcal{N}}$ be a Nash equilibrium strategy profile, and $\{d_i^*\}_{i \in \mathcal{N}}$ be corresponding critical node-sink separations at which nodes stop transmitting. Let also $d_{\max} = \max_{i \in \mathcal{N}} d_i^*$. Then, the following hold:*

- *If $d_{\max} < R$, then $g_i(d_i^*) = \frac{c_i}{1+c_i}$ for all $i \in \mathcal{N}$.*

- *If $d_{\max} = R$, then $g_i(d_i^*) = \frac{c_i}{1+c_i}$ for all $i \in \mathcal{N} - \{i_{\max}\}$ and $g_{i_{\max}}(R) \geq \frac{c_{i_{\max}}}{1+c_{i_{\max}}}$, where $i_{\max}$ is the index of the node whose critical threshold is equal to $R$.*

PROOF. We first prove the first item. When $d_{\max} < R$, all $d_i^*$'s belong to $(0, R)$. By using Theorem 1, we have $\mathbb{E}[u_i|T](d_i^*) = 0$ for all $i \in \mathcal{N}$. Since $\mathbb{E}[u_i|T](d) = (1 + c_i)g_i(d) - c_i$, the proof of the first item is finished.

To prove the second item, we first observe that $i_{\max}$ is well-defined, and there is only one node having its critical threshold equal to $R$ by Theorem 2. Thus, $d_i^* < R$ for all $i \in \mathcal{N} - \{i_{\max}\}$. This further implies that $g_i(d_i^*) = \frac{c_i}{1+c_i}$ for all $i \in \mathcal{N} - \{i_{\max}\}$ by Theorem 1. If a node with index $i_{\max}$ has its critical threshold equal to $R$, then $\mathbb{E}[u_{i_{\max}}|T](R) \geq 0$. Otherwise, $\mathbb{E}[u_{i_{\max}}|T](d)$ must hit zero for some $d < R$, which contradicts $i_{\max}$ having its critical threshold equal to $R$. Since $\mathbb{E}[u_{i_{\max}}|T](R) \geq 0$, we also have $g_{i_{\max}}(R) \geq \frac{c_{i_{\max}}}{1+c_{i_{\max}}}$. □

In the next theorem, we prove the equivalence of best-response strategies of nodes having the same cost of packet failures. This will help us to divide nodes into equivalence classes according to their costs of packet failures.

THEOREM 4. *At a Nash equilibrium, there cannot be two different nodes $i$ and $j$ having $c_i = c_j$ and $d_i^* \neq d_j^*$.*

PROOF. Suppose $\mathbf{s}^* = (s_i^*)_{i \in \mathcal{N}}$ is a Nash equilibrium strategy profile, and there are two different nodes $i$ and $j$ having $c_i = c_j$ and $d_i^* \neq d_j^*$. Without loss of generality, assume that $d_i^* < d_j^* \leq R$. Then, $g_i(d) = g_j(d)$ for all $d < d_i^*$ since the strategies of other nodes are fixed, and both of nodes $i$ and $j$ transmit for all $d < d_i^*$. Furthermore, $g_i(d_i^*) = g_j(d_i^*)$ by continuity of $g_i(d)$ and $g_j(d)$. We have $g_i(d_i^*) = \frac{c_i}{1+c_i}$ by Theorem 3. Since $c_i = c_j$, $g_i(d_i^*) = g_j(d_i^*)$ and $\mathbb{E}[u_j|T](d)$ is a non-increasing function of $d$, we have $\mathbb{E}[u_j|T](d) \leq 0$ for all $d \in [d_i^*, R]$, which contradicts $d_j^*$ being the least upper bound of node $j$'s best-response set. □

For any given cost vector $\mathbf{c} = (c_i)_{i \in \mathcal{N}}$, we divide nodes into equivalence classes as follows. For a given cost value $\tilde{c} > 0$, $\tilde{c}$-equivalence class of nodes is equal to

$$\tilde{\mathcal{C}} = \{i \in \mathcal{N} : c_i = \tilde{c}\}.$$

By Theorem 4, all nodes in $\tilde{\mathcal{C}}$ have the same strategies at a Nash equilibrium. We let $\tilde{\mathcal{N}} = \left\{ \tilde{\mathcal{C}}_1, \tilde{\mathcal{C}}_2, \cdots, \tilde{\mathcal{C}}_{|\tilde{\mathcal{N}}|} \right\}$ be the set of non-empty equivalence classes of nodes whose elements are ordered as

$$\tilde{c}_1 > \tilde{c}_2 > \cdots > \tilde{c}_{|\tilde{\mathcal{N}}|},$$

where $\tilde{\mathcal{C}}_i = \{j \in \mathcal{N} : c_j = \tilde{c}_i\}$.

In the next theorem, we will focus on the set of equivalence classes of nodes, and give a sufficient condition for a Nash equilibrium to exist.

THEOREM 5. *Let $\mathbf{s} = (s_i)_{i \in \mathcal{N}}$ be a monotonically decreasing strategy profile with cut-off points $(\tilde{d}_i)_{i \in \mathcal{N}}$, $\mathbf{c} = (c_i)_{i \in \mathcal{N}}$ be the cost vector of nodes, and $\tilde{\mathcal{N}} = \left\{ \tilde{\mathcal{C}}_1, \tilde{\mathcal{C}}_2, \cdots, \tilde{\mathcal{C}}_{|\tilde{\mathcal{N}}|} \right\}$ be the set of corresponding equivalence classes of nodes. Assume also that $\lambda \ll \mu$.*

1. $\left| \tilde{\mathcal{C}}_{|\tilde{\mathcal{N}}|} \right| > 1$: *For all $\tilde{\mathcal{C}}_i \in \tilde{\mathcal{N}}$, if there exists $\tilde{d}_i^*$ such that for all $j \in \tilde{\mathcal{C}}_i$, $d_j = \tilde{d}_i^*$ and $g_j(\tilde{d}_i^*) = \frac{\tilde{c}_i}{1+\tilde{c}_i}$, then $\mathbf{s}$ is a Nash equilibrium.*

2. $\left| \tilde{\mathcal{C}}_{|\tilde{\mathcal{N}}|} \right| = 1$: *For all $\tilde{\mathcal{C}}_i \in \tilde{\mathcal{N}}$, if there exists $\tilde{d}_i^*$ such that $\tilde{d}_{|\tilde{\mathcal{N}}|}^* = R$, and for all $j \in \tilde{\mathcal{C}}_i$, $1 \leq i \leq |\tilde{\mathcal{N}}| - 1$, $d_j = \tilde{d}_i^*$ and $g_j(\tilde{d}_i^*) = \frac{\tilde{c}_i}{1+\tilde{c}_i}$, then $\mathbf{s}$ is a Nash equilibrium.*

PROOF. We start the proof with the first item. Assume that the conditions in the first item hold. We then show that $\tilde{d}_1^* < \tilde{d}_2^* < \cdots < \tilde{d}_{|\tilde{\mathcal{N}}|}^*$. To this end, assume there exists $\tilde{\mathcal{C}}_i$ and $\tilde{\mathcal{C}}_k$ having $\tilde{c}_i > \tilde{c}_k$ and $\tilde{d}_i^* \geq \tilde{d}_k^*$. Take $j_i \in \tilde{\mathcal{C}}_i$ and $j_k \in \tilde{\mathcal{C}}_k$. By using the same arguments as in Theorem 4, we have

$$g_{j_i}(\tilde{d}_k^*) = g_{j_k}(\tilde{d}_k^*) = \frac{\tilde{c}_k}{1+\tilde{c}_k} < \frac{\tilde{c}_i}{1+\tilde{c}_i}.$$

By Property 2, we have $g_{j_i}(\tilde{d}_i^*) < \frac{\tilde{c}_i}{1+\tilde{c}_i}$, which is a contradiction.

Now, assume that there exists a $\tilde{\mathcal{C}}_i$ and a node $j \in \tilde{\mathcal{C}}_i$ such that $s_j(d) = 1$ for all $d \in [0, \tilde{d}_i^*)$, and 0 otherwise is not the best response of node $j$ given $\mathbf{s}_{-j}$. This implies that $d_j^* \neq \tilde{d}_i^*$. First consider $d_j^* < \tilde{d}_i^*$. By Property 3, $g_j(\cdot)$ is a strictly decreasing function on $(d_j^*, \tilde{d}_i^*)$. Thus, $\mathbb{E}[u_j|T](d_j^*) >$ 0, which is a contradiction. Now, consider $d_j^* > \tilde{d}_i^*$. If $i = |\tilde{\mathcal{N}}|$, then $\mathbb{E}[u_j|T](d) = 0$ for all $d \geq \tilde{d}_i^*$. This contradicts $d_j^*$ being the least upper bound for $\mathcal{B}_j$. If $i < |\tilde{\mathcal{N}}|$, then $\mathbb{E}[u_j|T](d_j^*) < 0$, which is another contradiction. Thus, $d_j^* = \tilde{d}_i^*$.

For the second item, we repeat the same arguments for all $i \leq |\tilde{\mathcal{N}}| - 1$. For the equivalence class $\tilde{\mathcal{C}}_{|\tilde{\mathcal{N}}|}$, let $j$ be the node in this class. Then, $g_j\left(\tilde{d}_{|\tilde{\mathcal{N}}|-1}^*\right) = \frac{\tilde{c}_{|\tilde{\mathcal{N}}|-1}}{1+\tilde{c}_{|\tilde{\mathcal{N}}|-1}} > \frac{c_j}{1+c_j}$. Since there is no node but node $j$ transmitting when node-sink separation is greater than $\tilde{d}_{|\tilde{\mathcal{N}}|-1}^*$, we have $\mathbb{E}[u_j|T](d) > 0$ for all $d \in [0, R]$. Therefore, $d_j^* = R$. □

## 3.4 Example: Uniformly Distributed Nodes Having the Same Cost of Packet Failures

To illustrate the use of the above theorems, we will now calculate the Nash equilibrium of the random access game with imperfect location information when $n$ nodes are uniformly distributed over the network domain. We assume that all nodes have the same cost of packet failures which is equal to $c > 0$. By Theorem 4, all nodes have the same strategies at Nash equilibrium. Therefore, we focus our attention on symmetric monotonically decreasing strategy profiles. Let $d^*$ be the critical node-sink separation such that nodes transmit when their distance to the sink node is smaller than $d^*$, and back off otherwise. By Theorem 2, we have $d^* < R$. Therefore, by Theorem 3, we must have

$$g_i(d^*) = \frac{c}{1+c}$$

for all $i \in \mathcal{N}$.

Since nodes have symmetric strategies, $g_i(d)$ will be the same for all $i \in \mathcal{N}$. Let us consider $g_1(d)$. For all $d < d^*$, we have

$$g_1(d) = \left(1 - \left(\frac{d}{R}\right)^2\right)^{n-1}.$$

By Property 4, we also have

$$g_1(d^*) = \left(1 - \left(\frac{d^*}{R}\right)^2\right)^{n-1}. \quad (5)$$

Therefore,

$$d^* = R \cdot \sqrt{1 - \left(\frac{c}{1+c}\right)^{\frac{1}{n-1}}}. \quad (6)$$

In Fig. 3, we show the change of $d^*$ as a function of $c$ for $R = 12$ and for different values of number of nodes. $d^*$ decreases as $c$ increases since nodes are deterred from transmission due to the increasing cost of packet failures. $d^*$ also decreases as $n$ increases due to the increasing number of nodes competing for communicating with the sink node.

## 4. EXTENSIONS AND FUTURE WORK

The current work suggests two interesting future research directions. The first one is to analyze ad-hoc networks in which there are many transmitter-receiver pairs. The behavior of selfish transmitters in this paper has been analyzed only for a *many-to-one* communication scenario where there are many transmitter nodes wishing to communicate with a common sink node. However, the set-up of the problem can be easily generalized to the communication scenario where there are *many one-to-one* communication links.

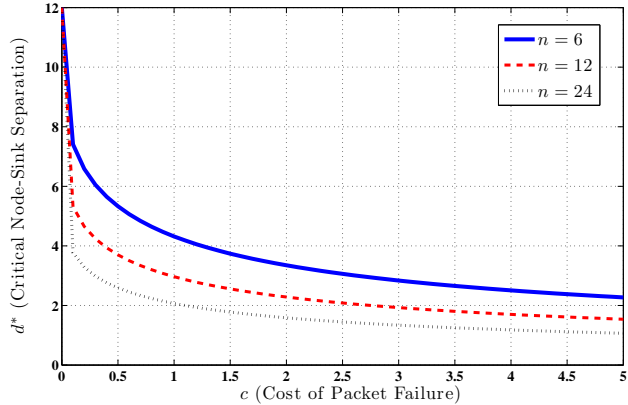

**Figure 3: Change of critical node-sink separation $d^*$ at which nodes stop transmitting as a function of the cost of packet failures $c$. ($R = 12$)**

To this end, consider a wireless ad-hoc network in which there are $n$ transmitter-receiver pairs. The distance between the $i$th pair is represented by $D_i$, where $D_i$ is a random variable whose distribution is known by other nodes but its exact value is not known. The transmitter and the receiver of the $i$th pair can be thought as if they are joined together by a stick of length $D_i$, and then thrown over the network domain by preserving $D_i$. The center of the stick joining the $i$th pair is distributed over the network domain according to a known spatial point process. Then, the $i$th pair communicates successfully if there is no transmitting node inside the disc of radius $D_i$ centered at the receiver of this pair, which is closely related to the problem analyzed in this paper. Therefore, a similar analysis reveals the Nash equilibrium strategy profiles for nodes in this more ad-hoc setting.

The second extension is to consider more general wireless channel models. In this work, we have considered only the wireless channel model in which the signal with the highest received power can be captured successfully. However, the same analysis can be repeated for more general multi-packet reception channel matrices where it is possible to simultaneously receive more than one packet successfully.

## 5. CONCLUSIONS

In this paper, we have analyzed the behavior of selfish transmitters under imperfect location information. To the best of our knowledge, this is the first work analyzing wireless networks containing selfish nodes under imperfect location information. The current work also opens some interesting future research directions, and provides insights into how to analyze multi-hop wireless ad hoc networks consisting of selfish nodes with multi-packet reception capabilities.

Our results can be summarized as follows. We first obtained the properties of packet success probability as a function of node-sink separation for arbitrary strategy profiles. We then characterized the behavior of best-response strategies of nodes against arbitrary opponent strategy profiles. In particular, we proved that the best-response strategy of a node against any arbitrary opponent strategy profile is a monotonically decreasing strategy in which the node transmits until a critical node-sink separation, and then stops transmitting. By using these results, we finally gave necessary and sufficient conditions for a strategy profile to be Nash equilibrium.